\documentclass[runningheads]{llncs}
	\usepackage{lmodern}
\usepackage[T1]{fontenc}
\usepackage{graphicx}
\usepackage{cite}
\usepackage{amsfonts}
\usepackage{textcomp}
\usepackage{xspace}
\usepackage{xcolor}
\usepackage{eso-pic}
\usepackage{tcolorbox}

\definecolor{preprintcolor}{RGB}{0,84,135}

\AddToShipoutPictureFG{%
  \AtPageUpperLeft{%
    \raisebox{-1.1cm}{%
      \makebox[\paperwidth][c]{%
        \tcbox[
          on line,
          arc=2pt,
          boxrule=0.4pt,
          colframe=preprintcolor,
          colback=preprintcolor!6,
          coltext=preprintcolor,
          left=5pt, right=5pt, top=2pt, bottom=2pt,
          fontupper=\footnotesize\sffamily
        ]{%
          \begin{tabular}{@{}c@{}}
            \textbf{Preprint} accepted for publication at the\\
            28th International Conference on Information Integration and Web Intelligence (iiWAS2026)
          \end{tabular}%
        }%
      }%
    }%
  }%
}

\usepackage{microtype}
\usepackage{hyperref}
\usepackage{url}
\usepackage{breakurl}

\usepackage{csquotes}
\usepackage{amssymb}
\usepackage{booktabs}
\usepackage{array}
\usepackage{etoolbox}
\AtBeginEnvironment{thebibliography}{\setlength{\itemsep}{-3pt}\setlength{\parsep}{0pt}}

\begin{document}
	\title{FireOPS: A Human-Centric Emergency Operations Platform}
	\titlerunning{FireOPS: A Human-Centric Emergency Operations Platform}
	%
	%
	\author{Alexander Wallner\inst{1,2}\orcidID{0009-0001-4489-1859} \and
		Michael Riegler\inst{1,2}\orcidID{0000-0002-6937-986X} \and
		Philipp Neuhauser\inst{1,3}\orcidID{0009-0000-8756-8181} \and
		Helmut Haider\inst{1,3} \and
		Lukas Neuhauser\inst{1,3} \and
		Marco Stadler\inst{2}\orcidID{0000-0001-8790-6048} \and
		Eduard Paireder\inst{1,4}}
	\authorrunning{Wallner et al.}
	%
	\institute{District Fire Command Perg - Team FireOPS, Austria\\
		\email{contact@fireops.at}\\
		\url{https://pe.ooelfv.at/fireops} 
		\and
		Johannes Kepler University Linz, Linz, Austria\\
		\and
		University of Applied Sciences Upper Austria, Hagenberg, Austria\\
		\and
		Upper Austrian Fire Brigade Association (OÖLFV)\\
	}
	\maketitle              
	\begin{abstract}
	Volunteer fire brigades rely on fragmented alerting channels and radio plus paper for situational awareness, hindering a shared common operational picture (COP). FireOPS integrates smartphone alerting, navigation, a live geospatial COP, a district resource-load overview and contextual manuals into a human-centric decision-support surface for commanders and responders, running on an edge + cloud architecture whose dual-channel provides source-level redundancy against failure of either the radio or the provincial server feed; delivery to smartphones still depends on the cloud core and mobile networks. This lessons-learned report presents the design rationale, positions FireOPS within emergency-response information systems research, and reports operational findings from a primary log extract of a major regional flood in 2024 and from the 2025 public app release. A feature-matrix comparison with DIVERA~24/7, blaulichtSMS and the LifeRescue prototype identifies COP, decision support and dual-ingestion redundancy as the main differentiators, and we map the operational gaps observed during the flood to concrete engineering responses, deployed or in active development.
		
		\keywords{Common operational picture \and Situational awareness \and Crisis informatics \and Volunteer fire brigade \and Information integration \and Edge + Cloud computing \and Web services integration \and Human-centric design}
	\end{abstract}
	%
	%
	%
	\section{Introduction}
	\label{sec:intro}
	
	Time-critical emergency response relies on personnel and commanders building and maintaining an accurate model of the evolving situation. In Endsley's three-stage theory of situation awareness---perception, comprehension and projection---anticipating future states is crucial for high-quality decision-making~\cite{endsley1995}. In multi-agency operations, information must cross organisational boundaries to form a \emph{common operational picture} (COP), and crisis-management research shows that COPs support collective sense-making when they structure problems and let responders negotiate meaning rather than simply displaying more data~\cite{wolbers2013}. Recent work on human-centred crisis informatics identifies mobile applications, bidirectional warnings and responder-side dashboards as priority directions~\cite{kaufhold2024}.
	
	Volunteer fire brigades are staffed not by career professionals but by unpaid community members responding to alarms on an on-call basis, a distinctive context for information integration: members may be off-site when an alarm arrives, have limited training on specialised IT systems, and depend on community resources. Fragmentation across dispatch apps, paper maps, radio communications and separate reference manuals introduces delays and increases cognitive load. The FireOPS project arose from these challenges in the Upper Austrian district of Perg. Its central objective was to give incident commanders and responders a \emph{shared, district-wide COP} unifying, in a single platform, a live map of incidents, dispatched units, vehicles, hydrants and weather; a district-load overview that surfaces municipal hotspots during peak events; mobile alerting with explicit \enquote{coming\,/\,not coming} acknowledgment; one-click navigation; and context-specific manuals and rescue cards. Together these layers constitute an information-integration problem at the intersection of three topic areas: IoT and edge computing (per-brigade paging receivers, on-premise brokers, brigade-information displays); web services and API integration (REST Core~API, operational-management platform synchronisation, mobile push notifications, and a redundant feed from Austria’s fire-service \textit{Warn- und Alarmsystem} (WAS) radio alerting network~\cite{bmi_warnung_alarmierung}); and large language models for web intelligence (an assistant with retrieval-augmented generation over a fire-service corpus). Heterogeneous sources must be fused into a single human-centric decision-support surface usable by volunteer practitioners under real-time load.
	
	Against this objective, the paper reports three contributions and two supporting artefacts: $(i)$~an end-to-end edge\,+\,cloud architecture with a dual-channel (radio\,+\,server) redundancy design (Sect.~\ref{sec:architecture}); $(ii)$~a dataset (primary log evidence) from a major regional flood event in September~2024 comprising 233~incidents and 630~dispatch records, with a descriptive response-time Cumulative Distribution Function (CDF) which, to the best of our knowledge, is the first published response-time baseline from an operational volunteer-fire-brigade deployment (Sect.~\ref{sec:log_evidence}); and $(iii)$~a structured \emph{gap-to-response} reflection mapping each missing capability during the 2024 flood to a corresponding engineering fix (deployed or in active development), triangulated against 204~in-app feedback entries and 31~email threads (Sect.~\ref{sec:feedback} \& Sect.~\ref{sec:reflection}). Supporting artefacts are a brigade-level adoption analysis over 29~months and 132~installations (Sect.~\ref{sec:adoption}) and a feature-matrix comparison against DIVERA~24/7, blaulichtSMS and the LifeRescue research prototype (Sect.~\ref{sec:comparison}). The paper is a \emph{lessons-learned report} rather than a controlled evaluation: no quantitative claims about response-time improvements or comparative effectiveness are made. FireOPS is operated non-commercially by volunteer members of the Bezirks-Feuerwehrkommando Perg (District Fire Command Perg) and has not received commercial funding.
	
	\section{Related Work}
	\label{sec:related-work}
	
	\subsection{Shared situational awareness and common operational pictures}
	
	Situation awareness (SA) originates in aviation and has been adapted to emergency response; Endsley's framework defines it as the perception of elements in the environment, comprehension of their meaning and projection of their future status~\cite{endsley1995}. For emergency managers, SA must be distributed across individuals and agencies; Bharosa et al.\ characterise disaster information sharing as a socio-technical challenge shaped by organisational trust and communication protocols~\cite{bharosa2010}. Wolbers and Boersma caution that a COP is not automatically a common understanding: it must support collective sensemaking and problem representation, functioning as a \enquote{trading zone} across agencies~\cite{wolbers2013}. Steen-Tveit and Munkvold distinguish a COP from \emph{common situational understanding}, arguing that shared technology, mutual respect and trust are preconditions for the latter~\cite{steentveit2021}, and later derive a taxonomy of multi-agency information requirements that maps closely onto the layered design of responder-facing dashboards such as FireOPS~\cite{steentveit2023}. Recent reviews extend this lineage with human-centred crisis informatics and citizen integration in emergencies~\cite{kaufhold2024}.
	
	\subsection{Firefighter information systems and adoption}
	
	Firefighter-specific information systems research emphasises user-centred design and role-specific information delivery~\cite{weidinger2018, yang2009}; Kapalo and LaViola surveyed 50~firefighters internationally and found severe difficulty gathering, accessing and retrieving pre-incident planning information, pointing to interface design as a key obstacle~\cite{kapalo2019}. Adoption studies in voluntary fire brigades show that technical sophistication alone does not guarantee acceptance: usability, reliability and alignment with existing workflows are critical~\cite{weidinger2018,weidinger2024}. Closest in concept to FireOPS is LifeRescue, a web-based responder application from the Agder group combining mapping, incident overview and resource views in one browser-based tool~\cite{nunavath2019}.
	
	\subsection{Geospatial decision support and wildfire DSS}
	
	Wildfire decision-support systems such as the U.S. Wildland Fire Decision Support System (WFDSS)~\cite{noonanwright2011} and the Greek regional prototype Virtual Fire~\cite{kalabokidis2013} integrate weather data, fire-spread models and strategic decision support. They share with FireOPS the premise that geospatial data is the substrate on which situational awareness is built: hazard locations, resource positions and terrain are fused into map layers on which commanders reason spatially about an evolving event. The FireOPS COP is assembled the same way, from incident coordinates, dispatched units and vehicles, hydrants and water points, and weather, exchanged as GeoJSON layers so that everyone works from one shared map. The difference is scale and purpose: where wildfire DSS operate strategically and model fire spread over hours to days, FireOPS applies the principle tactically and in real time, integrating routing, alerting and reference materials for the all-hazard incident mix that dominates volunteer service.
	
	\subsection{Interoperability standards and population-facing alerting}
	
	The OASIS Common Alerting Protocol (CAP), a content standard for alert messages~\cite{oasis_cap}, enables dissemination across heterogeneous platforms, while the OGC SensorThings API~\cite{ogc_sensorthings} and OGC API Features~\cite{iso19168} support web-based sharing of sensor observations and geospatial features. Population-facing warning applications in the DACH region such as NINA, KATWARN and AT-Alert address the public rather than responders; several build on CAP, others on cell-broadcast channels such as EU-Alert. FireOPS currently uses proprietary interfaces---the syBOS~\cite{solarys_sybos} HTTP API (the de facto Austrian fire-service operational-management platform) for personnel and vehicle synchronisation, and the provincial WAS feed for alerting---so these standards are pertinent to future integration with regional dashboards and weather feeds.
	
	\section{System Description}
	\label{sec:system-description}
	
	\subsection{Origin and practitioner-led design}
	
	FireOPS was conceived and prototyped by actively serving fire-brigade members in collaboration with the District Fire Command Perg, with the explicit aim of supporting fire-service operations end-to-end~\cite{fireops_pe_ooelfv}. The authors are active members of different volunteer brigades within district Perg; design decisions are informed both by their own operational experience and by structured feedback from non-author brigades and from the District Fire Commander. Onward adoption into neighbouring districts was driven by peer brigades promoting the platform from their own deployment experience rather than by the author team, which we read as partial external validation of the practitioner-led design; the residual conflict of interest is set out in the disclosure statement below. This origin matches the acceptance factors identified in the firefighter-IT adoption literature~\cite{weidinger2024,weidinger2018}: alignment with existing workflows and the perception that the system is built \emph{for} responders rather than \emph{at} them are recurring themes in the public coverage~\cite{umweltpreis2024}. During the pilot phase the platform was exposed to a diverse incident mix---traffic accidents, building and vegetation fires and the September~2024 flood---stress-testing its information layers beyond any single hazard category~\cite{fireops_pe_ooelfv}. Subsequent recognition, including the 2024 Upper Austrian Environmental Prize~\cite{umweltpreis2024}, appears to have lowered the political cost of onward deployment, although prize recognition is not evidence of operational effectiveness.
	
	\subsection{Architecture overview}
	\label{sec:architecture}
	
\begin{figure}[tb]
		\centering
		\includegraphics[width=0.70\textwidth]{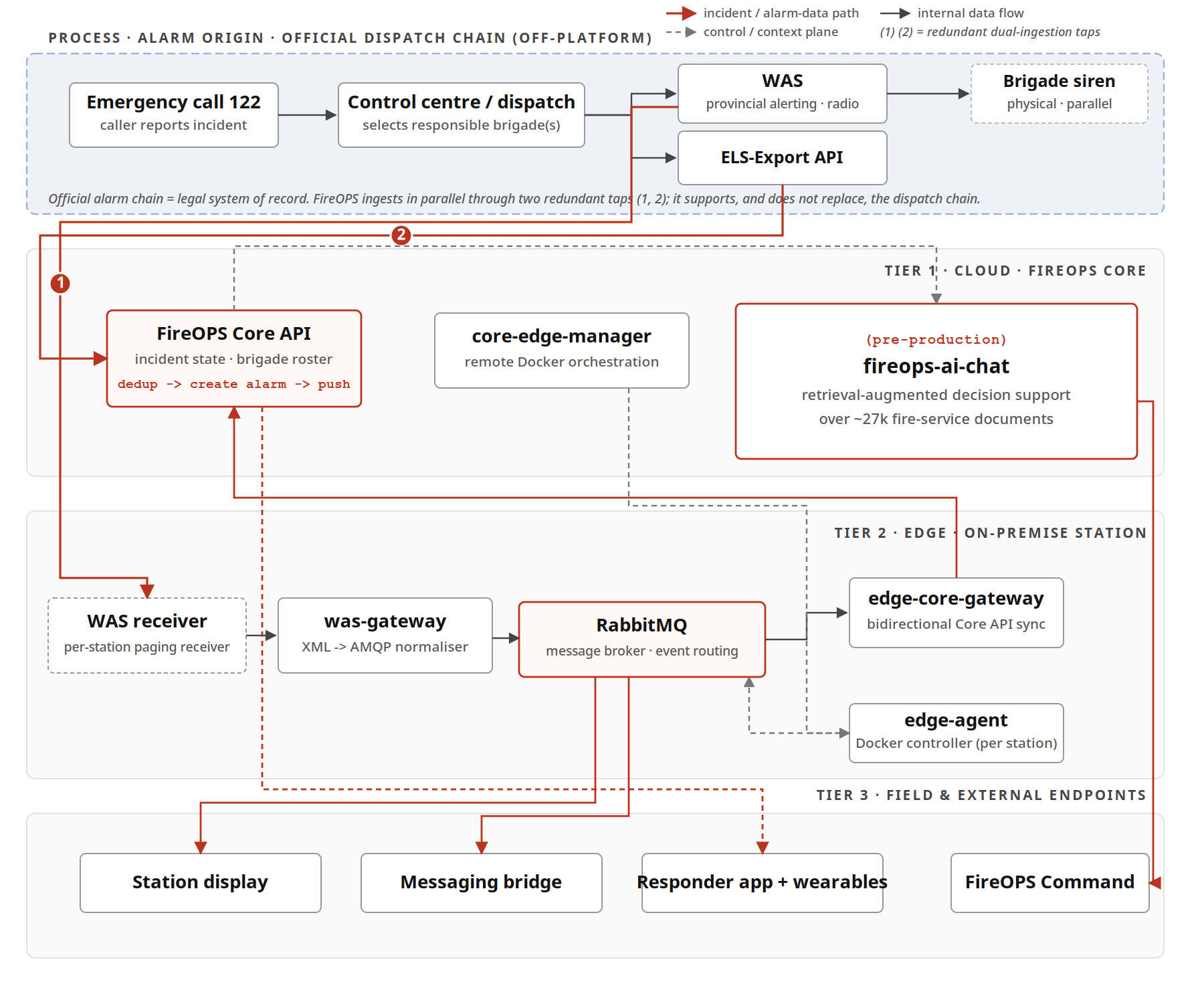}
		\caption{Component-level architecture across three tiers. 
		}
		\label{fig:architecture}
	\end{figure}

	The platform exposes five information layers: alerting with responder acknowledgement, incident information with one-click navigation, a geospatial COP (incidents, responding units, vehicles, weather, hydrants), a district-wide resource-load overview that highlights municipal \enquote{hotspots} during high-load events, and a context-specific knowledge layer surfacing rescue cards, hazardous-materials sheets and operating manuals. The deployed architecture is partitioned into three tiers (Fig.~\ref{fig:architecture}). At the \emph{cloud / core} tier, a REST Core~API (\texttt{fireops.at}) holds authoritative incident state and brigade rosters; a Go control-plane service (\texttt{core-edge-manager}) supervises per-station Docker workloads over WebSocket; and a pre-production retrieval-augmented assistant (\texttt{fireops-ai-chat}) provides decision support over an index of $\approx$\,27\,000 fire-service documents (operational manuals, rescue cards, hazardous-materials sheets and service regulations).
	At the \emph{edge} tier, each fire station runs an on-premise stack: a per-station WAS paging receiver~\cite{strabag_alu2g} feeds XML over TCP into a \texttt{was-gateway} which normalises events onto a local message broker (RabbitMQ~\cite{rabbitmq}; three exchanges: \texttt{events}, \texttt{units}, \texttt{health}). The broker routes messages to the station dashboard and to online messaging services (Telegram, WhatsApp and similar), while an \texttt{edge-core-gateway} bidirectionally synchronises state with the Core~API. Because every event is published on the local broker, further consumers---additional messaging bridges, home-automation actuators, signage---can be attached without modifying the core. The \emph{field} tier surfaces information through the station-house display, the messaging-service channels, responder smartphones (push with accept\,/\,decline acknowledgement) and FireOPS~Command, the commander's web interface; Apple~Watch and Wear~OS companions and Android~Auto and Apple~CarPlay in-vehicle clients extend alerting and acknowledgement to the wrist and the vehicle.
	
	An alarm originates in the statutory dispatch chain, onto which FireOPS layers a \emph{supporting} system rather than a replacement. An emergency call reaches the public-safety answering point (Euro-emergency number 112 or the Austrian fire-brigade number 122), where a dispatcher identifies the responsible brigade and triggers it over radio through the provincial alerting infrastructure (WAS); the same trigger drives the station's siren. FireOPS taps this chain through two redundant ingestion paths. Path~\textcircled{1} is a radio uplink: the WAS page is received on the per-station receiver and normalised by the \texttt{was-gateway}. Path~\textcircled{2} is a server-side feed: the dispatcher's entry also propagates to the provincial ELS-Export interface, which the Core~API polls at a 10\,s cadence. The two streams are deduplicated at the backend, which creates the incident and emits the responder push, bounding end-to-end alerting latency by the faster channel and degrading gracefully under partial failure of either.
	
	Two choices warrant emphasis. First, the edge\,+\,cloud partition isolates each station against transient cloud loss: while disconnected, the station's RabbitMQ broker keeps delivering alarms to the local dashboard and messaging-service channels, and the \texttt{edge-core-gateway} reconciles state once connectivity returns. Second, dual-path ingestion provides source-level redundancy so that no single link, radio receiver or provincial server is a single point of failure; this is not hypothetical, as during regional peak load the server feed degrades while the radio-side WAS path keeps delivering pages. Personnel and vehicle records are synchronised via the syBOS API~\cite{solarys_sybos}.
	
	\subsection{Adoption trajectory}
	\label{sec:adoption}
	
\begin{figure}[tb]
		\centering
		\includegraphics[width=0.65\textwidth]{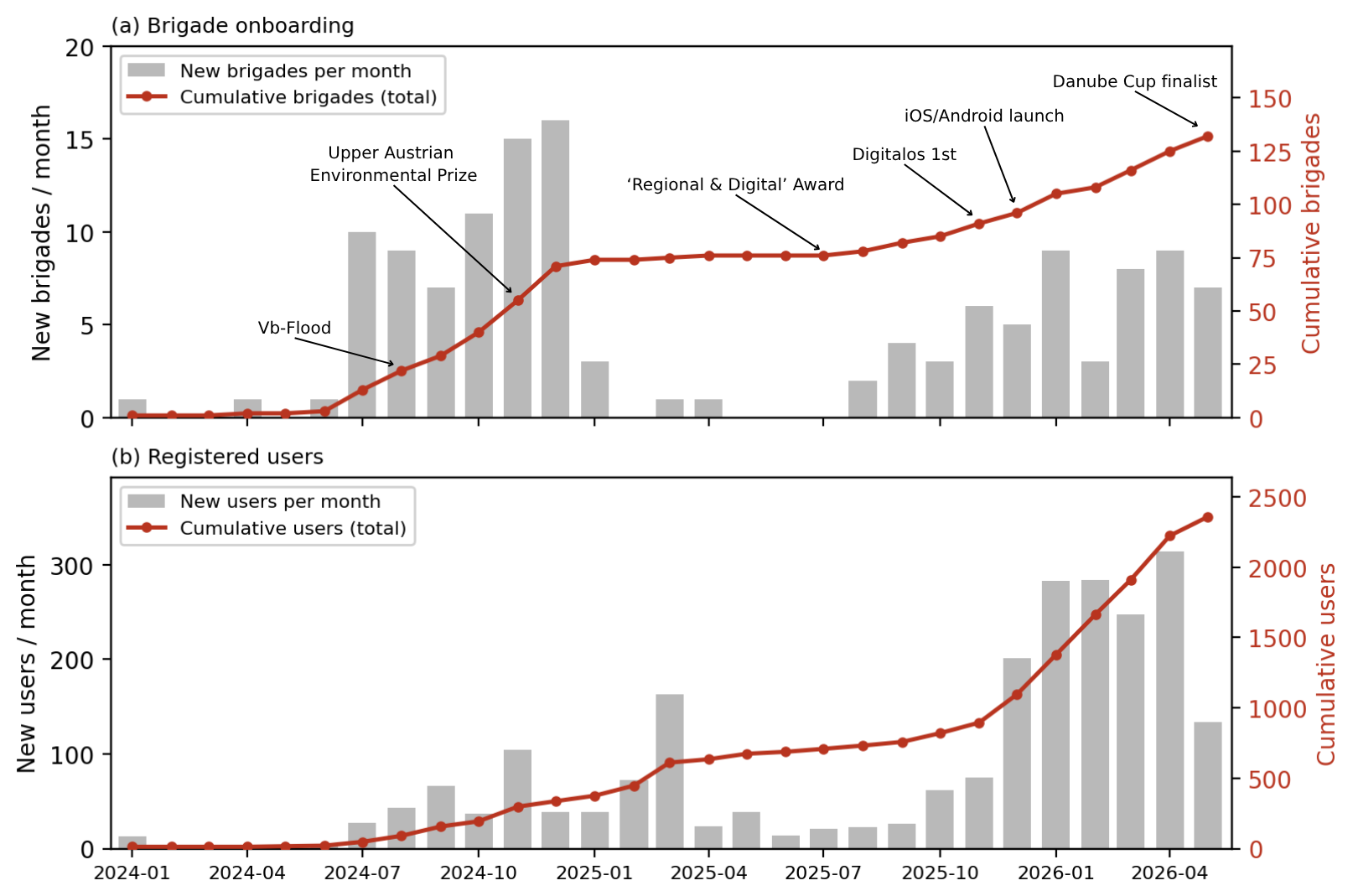}
		\caption{FireOPS adoption from January~2024 to May~2026 on a shared monthly axis. Bars (left axis) are new (a)~brigades and (b)~users per month; lines (right axis) are the cumulative totals. The user surge follows the public iOS and Android launch on 24~December~2025. Source: FireOPS internal registry.
		} 
		\label{fig:adoption}
	\end{figure}

	Fig.~\ref{fig:adoption} (a) plots the adoption curve at brigade granularity, derived from the platform's internal registry. The May~2026 installed base of 132 brigades is roughly 15\,\% of the 906 brigades in Upper Austria and 3\,\% of the 4{,}767 nationwide~\cite{oebfv2024}, concentrated in Perg and adjacent OÖLFV districts. Four regimes are visible: (i)~a pilot phase (January--June 2024) with single-digit installations at the District Fire Command Perg and partner brigades; (ii)~a first growth wave (July--December 2024) with 7--16 new brigades per month taking the installed base from 3 to 71, preceding and spanning with the September~2024 Vb-cyclone flood and the November~2024 Upper Austrian Environmental Prize~\cite{umweltpreis2024}; (iii)~a plateau (January--August 2025) of 0--3 new brigades per month while the platform remained formally a pilot with no app-store release; and
	(iv)~a second growth wave (September~2025 onward) re-accelerating the curve in temporal alignment with two further awards~\cite{regionalitaetspreis2025,digitalos2025}, the public iOS and Android app-store launch on 24~December~2025, and a Danube~Cup~2026 finalist nomination~\cite{danubecup2026}. The curve is descriptive, not causal: we claim no single event \emph{caused} the inflexions and do not control for confounders such as regional outreach or word-of-mouth. The alignment between recognition events, app-store availability and adoption is nonetheless consistent with public legitimacy lowering the perceived risk of adoption for downstream brigades. Panel~(b) overlays registered-user growth on the same axis: registrations remained in the hundreds throughout the pilot period, then climbed steeply after the launch, reaching 2{,}358 by May~2026. Public distribution thus widened individual reach far faster than institutional onboarding, which grew more steadily at brigade granularity.
	
	\subsection{Comparison to related platforms}
	\label{sec:comparison}
	
	Table~\ref{tab:feature_matrix} compares FireOPS with two established responder-facing alerting and availability platforms used in the Austrian/DACH emergency-services context, DIVERA~24/7~\cite{divera247} and blaulichtSMS~\cite{blaulichtsms}, and with the LifeRescue research prototype~\cite{nunavath2019}, included because it covers research-oriented features beyond those typically documented for commercial alerting platforms. The feature dimensions are derived from the FireOPS architecture and product documentation~\cite{fireops_pe_ooelfv} and from public product material for the commercial systems; cell markings follow the legend in Table~\ref{tab:feature_matrix}.
	
	\begin{table}[t]
		\centering
		\caption{Feature comparison between FireOPS and the most established peer systems for Austrian volunteer fire brigades. Legend: $\checkmark$ first-class, $\circ$ partial / indirect, -- absent / no evidence. \emph{Caveat:} FireOPS is assessed by the developing team with full internal knowledge; peer systems from public product material (May~2026)~\cite{divera247,blaulichtsms,nunavath2019}.}
		\label{tab:feature_matrix}
		\setlength{\tabcolsep}{4pt}
		\renewcommand{\arraystretch}{1.0}
		\footnotesize
		\begin{tabular}{|>{\raggedright\arraybackslash}p{5.2cm}|*{4}{>{\centering\arraybackslash}p{1.35cm}|}}
			\hline
			\textbf{Capability} & FireOPS & DIVERA 24/7 & blaulicht\-SMS & Life\-Rescue \\
			\hline
			
			\multicolumn{5}{|l|}{\emph{Alerting and acknowledgement}} \\
			\hline
			Mobile push alerting                                  & $\checkmark$ & $\checkmark$ & $\checkmark$ & -- \\
			SMS / voice / pager fallback channels                 & $\circ$           & $\checkmark$ & $\checkmark$ & -- \\
			``Coming / not coming'' acknowledgement               & $\checkmark$ & $\checkmark$ & $\checkmark$ & -- \\
			Geofence-based automatic availability                 & --           & $\checkmark$ & --           & -- \\
			Wearable companion (Apple Watch / Wear OS)            & $\circ$ & $\circ$      & --           & -- \\
			In-vehicle alerting (Android Auto / CarPlay)          & $\circ$ & --           & --           & -- \\
			\hline
			
			\multicolumn{5}{|l|}{\emph{Common operational picture}} \\
			\hline
			Incident map with hydrants and water points           & $\checkmark$ & $\circ$      & $\checkmark$ & $\circ$ \\
			One-click navigation to scene                         & $\checkmark$ & $\checkmark$ & $\checkmark$ & -- \\
			Live unit / vehicle status                            & $\checkmark$ & $\checkmark$ & --           & $\circ$ \\
			District-wide resource-load overview                  & $\checkmark$ & --           & --           & -- \\
			Station-house operations display                      & $\checkmark$ & $\circ$      & $\checkmark$           & -- \\
			\hline
			
			\multicolumn{5}{|l|}{\emph{Decision support and knowledge}} \\
			\hline
			Context-specific manuals / rescue cards               & $\checkmark$ & --           & --           & -- \\
			AI assistant over fire-service corpus (pre-production) & $\checkmark$ & -- & -- & -- \\
			\hline
			
			\multicolumn{5}{|l|}{\emph{Architecture and integration}} \\
			\hline
			Hardware paging integration (e.g.\ WAS, TETRA)        & $\checkmark$ & $\checkmark$ & $\circ$ & -- \\
			Edge\,+\,cloud hybrid architecture                    & $\checkmark$ & --           & --           & -- \\
			Parallel deduplicated ingestion (radio\,+\,server)    & $\checkmark$ & $\circ$      & $\circ$      & -- \\
			Device-level resilience (override, offline cache) & $\checkmark$ & $\circ$ & -- & -- \\
			Peer-reviewed evaluation evidence                     & --           & --           & --           & $\checkmark$ \\
			\hline
		\end{tabular}
	\end{table}
	
	Three observations follow. First, DIVERA~24/7 and blaulichtSMS are mature in the \emph{alerting} and \emph{availability} dimensions but offer little above the alerting channel: neither provides a district-wide resource-load overview or contextual manuals, and neither integrates an AI knowledge assistant. Second, FireOPS' principal differentiators sit in the COP and decision-support dimensions: a district-wide hotspot overview, a context-specific manual layer and a pre-production retrieval-augmented assistant grounded in the same corpus. Third, although FireOPS currently uses proprietary interfaces, its architecture is the only one in the set with a documented edge\,+\,cloud topology and an explicit dual-ingestion redundancy path; standards-based interoperability (CAP, OGC SensorThings) is planned, not deployed. Both commercial systems can ingest alarms over more than one channel (DIVERA~24/7 via dispatch-centre interfaces or an alarm-parsing server; blaulichtSMS via TETRA readout or API~\cite{divera247,blaulichtsms}), but neither documents the simultaneous, backend-deduplicated ingestion FireOPS runs, so this is an architectural refinement rather than a functional gap. LifeRescue~\cite{nunavath2019} is orthogonal: a 2016 research prototype with a peer-reviewed evaluation but no ongoing public product presence, so it is assessed from its original publication. Against regional wildfire DSS such as Virtual Fire~\cite{kalabokidis2013}, FireOPS lacks fire-spread modelling but provides a finer-grained, responder-centric tactical view suited to the all-hazard mix that dominates volunteer service.
	
	\section{Lessons Learned from Operational Deployment}
	\label{sec:lessons-learned}
	
	\subsection{Deployment context: the September 2024 Perg flood}
	
	In September 2024, a Vb-cyclone produced prolonged rainfall across Lower and Upper Austria, with the district of Perg among the most heavily affected areas~\cite{landooe2024}; a Vb (Van Bebber) track carries Mediterranean moisture northwards over Central Europe and is the synoptic pattern behind most major regional floods. Contemporaneous reporting documented 250 fire-brigade operations and approximately 1{,}600 emergency personnel deployed in the district on 15~September 2024~\cite{orf2024}. The syBOS extract analysed here covers a different window and unit of analysis: 233~discrete incidents dispatched across the district between 14 and 17~September, of which 68 fall on 15~September. The dispatch log contains 630~records: 255~brigade-level alerts and 375~individual vehicle dispatches (mean 1.6~vehicles per incident, maximum 7; Fig.~\ref{fig:flood_evidence}(a)). FireOPS, then still a pilot, was operated by the district command centre and by the 27~local brigades that had voluntarily adopted it during its first eight months---a self-selected, technology-forward early-adopter subset rather than a representative sample; the log extract is district-wide rather than restricted to them. Exposure in this window also included a traffic accident at Klam and a major fire at Dimbach, documented publicly as the project's first use cases~\cite{fireops_pe_ooelfv}. The flood is the high-load anchor of this analysis but not the platform's only exposure: between January~2024 and May~2026 the brigades running FireOPS handled several thousand incidents across all hazard classes, each debrief feeding the lessons learned loop in Sect.~\ref{sec:reflection}.
	
	\subsection{Sources and scope}
	
	Observations and quantitative claims in this section draw on four sources: $(i)$~public project documentation~\cite{fireops_pe_ooelfv}; $(ii)$~Austrian press coverage of the flood and of subsequent project recognition~\cite{orf2024,umweltpreis2024,digitalos2025,regionalitaetspreis2025,danubecup2026}; $(iii)$~recurring practitioner statements by project leads and the District Fire Commander reproduced in those sources; and $(iv)$~a primary log extract for the four-day flood window (233~incidents, 630~dispatch records). One structural caveat: at flood time the responder-side acknowledgement function did not yet exist, so the syBOS~\cite{solarys_sybos} log records brigade-level dispatch acknowledgement and vehicle-level en-route/on-scene timestamps but no per-responder acknowledgement, the quantity the FireOPS acknowledgement layer now captures. That layer (Sect.~\ref{sec:architecture}) closed the gap; claims about acknowledgement rates are therefore deferred to the planned evaluation outlined in Sect.~\ref{sec:future_work}.
	
	\subsection{Primary log evidence from the syBOS extract}
	\label{sec:log_evidence}
	
\begin{figure}[tb]
		\centering
		\includegraphics[width=0.66\textwidth]{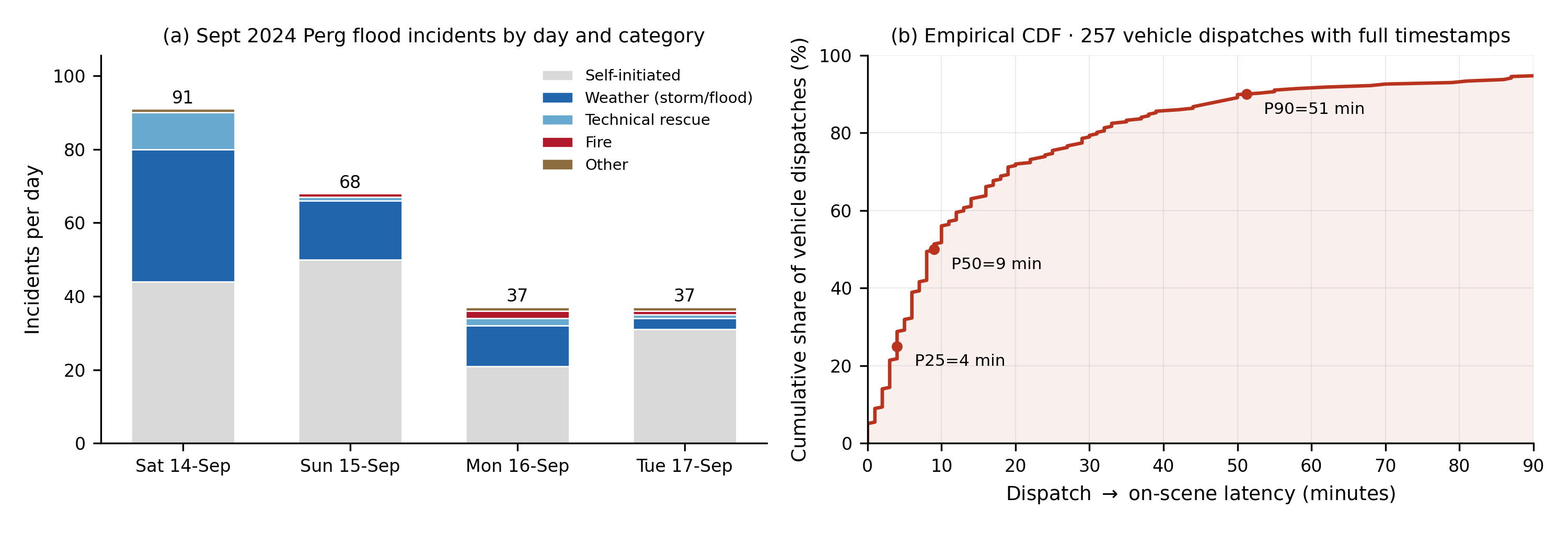}
		\caption{Primary log evidence for the September~2024 Perg flood. (a)~Daily incident counts by category over the four peak days, totalling 233~incidents. (b)~Empirical CDF of dispatch~$\to$~on-scene latency for the 257~vehicle dispatches with complete timestamps; markers are the 25\textsuperscript{th}, 50\textsuperscript{th} and 90\textsuperscript{th} percentiles. The x-axis is truncated at 90~min; 5\,\% of dispatches exceed it (maximum 539~min). Results are at vehicle, not individual responder, level.}
		\label{fig:flood_evidence}
	\end{figure}

	Fig.~\ref{fig:flood_evidence}(a) shows the daily incident workload across the four peak days. The district handled 91~incidents on 14~September, 68 on 15~September and 37 on each of 16 and 17~September; the workload comprised 146~self-initiated brigade operations, 66~severe-weather incidents, 14~technical-rescue incidents, 4~fire alarms and 3~other incidents. Within the weather-related group, flood-protection duties (28) and water damage (18) dominated. The load is spread across 26~municipalities (most affected: Baumgartenberg~22, Mauthausen~20, Luftenberg~19, Perg~town~19, Waldhausen~17, Schwertberg~15, Grein~14). Of the 226~incidents with vehicle-level records, 95 drew more than one vehicle (up to 7) and 11 drew vehicles from more than one brigade---the regime in which a shared COP is most valuable.
	
	Fig.~\ref{fig:flood_evidence}(b) shows the empirical CDF of dispatch~$\to$~on-scene latency for the 257~vehicle dispatches with both timestamps present and a non-negative interval. The distribution is right-skewed ($P_{25}=4$~min [3,~5]; $P_{50}=9$~min [8,~11]; 95\,\% bootstrap confidence intervals, $B=5{,}000$), consistent with short intra-municipal travel, with an extended right tail ($P_{90}=51$~min [41,~81]) for mutual-aid units traversing the wider district under degraded road conditions. These are descriptive baselines, not effectiveness claims, and serve as a reference for the controlled evaluation planned in Sect.~\ref{sec:future_work}.
	
	\paragraph{Missing-data caveat.}
	The 630~log records are of two structurally different kinds. The 255~brigade-level rows carry \emph{dispatch} and \emph{accepted} timestamps but never en-route or on-scene ones; the 375~vehicle rows carry \emph{en-route} for every row and \emph{on-scene} for 262 (70\,\%). The relevant denominator for the latency analysis is therefore the 375~vehicle dispatches, not all 630~records. On-scene is populated only when the crew pushes the corresponding status code from the vehicle terminal into the digital-radio backend; status messaging had been rolled out in Upper Austria only shortly before September~2024 and was not yet routine, and a multi-day high-tempo response further depresses compliance. Of the 262~dispatches carrying both timestamps, 5 yield a negative interval and are excluded, leaving 257; the 13~zero-minute values are retained, as they plausibly represent units already at or adjacent to the scene rather than data errors. En-route timestamps are deliberately not analysed: they equal the dispatch timestamp exactly for 373 of 375~vehicle rows, indicating the field is copied from dispatch rather than recorded from a status push, so this extract cannot speak to alerting-pipeline latency. The missing on-scene data is not completely at random (brigades less familiar with status messaging plausibly differ), so we treat the CDF as a conservative descriptive baseline and avoid inferential claims.
	
	\subsection{Practitioner feedback: triangulating the reflection}
	\label{sec:feedback}
	
	Beyond the syBOS log and the public coverage, the project maintains an in-app feedback form and a public contact channel. Between August~2024 and May~2026 these collected 204~in-app entries from 29~distinct users (108~ideas, 81~bug reports, 12~support requests, 3~uncategorised) and 31~longer email threads from brigade IT contacts and commanders. Coded against Table~\ref{tab:reflection}, the largest clusters concern the COP and live vehicle status (79~entries), responder acknowledgement (45), the station-house display (10) and knowledge-layer features, including requests for retrieval-augmented support grounded in the operational manuals (11). Eight of the 31~email threads come from brigades outside Perg (Steyr-Land, Linz-Land, Urfahr, Wels-Land, Freistadt, Braunau, Vorarlberg), consistent with the cross-district curve in Fig.~\ref{fig:adoption}.
	
	Recurring themes are geofence-based acknowledgement, iOS critical-alert override, drone-imagery integration, and migration friction from other systems like blaulichtSMS. The principal limitation is selection bias (in-app authors skew technically engaged, email correspondents self-selected), so a stratified follow-up survey is part of the evaluation planned in Sect.~\ref{sec:future_work}.
	
	\subsection{Gap-to-response reflection}
	\label{sec:reflection}
	
	In the September~2024 flood the layer that mattered most to commanders was the district-wide COP: a single map at the command centre showing every active incident, every dispatched unit and the load per municipality, which the project team observed to be the platform's principal contribution to coordination. Almost everything else that today distinguishes FireOPS in Table~\ref{tab:feature_matrix} was missing during the flood. Table~\ref{tab:reflection} maps each gap to the lesson learned on-scene and to the engineering response shipped since.
	
	\begin{table}[t]
		\centering
		\caption{Operational gaps observed during the September~2024 FireOPS deployment and the engineering responses implemented or prepared by May~2026. The district-wide COP is excluded: it was already available in 2024 and practitioners identified it as the most useful flood-response capability.}
		\label{tab:reflection}
		\setlength{\tabcolsep}{3pt}
		\renewcommand{\arraystretch}{1.05}
		\scriptsize
		\begin{tabular}{>{\raggedright\arraybackslash}p{2.6cm} >{\raggedright\arraybackslash}p{4.45cm} >{\raggedright\arraybackslash}p{4.45cm}}
			\hline
			\textbf{Gap in Sept 2024} & \textbf{Observation on-scene} & \textbf{Engineering response (shipped or pre-production by May 2026)} \\
			\hline
			Responder-side acknowledgement
			& Commanders had to call station chiefs or wait for radio check-ins to estimate available staffing.
			& Native \enquote{coming\,/\,not coming} acknowledgement on the responder app, with live tallies on the station display and at the command centre. \\
			
			Public mobile distribution
			& Pilot-only distribution restricted FireOPS to insider brigades; cross-boundary mutual aid fell back to radio and phone, with the COP unavailable to incoming units.
			& Public release on iOS and Android on 24~December~2025; cumulative installed base grew from 27~brigades at flood time to 132 by May~2026 (Fig.~\ref{fig:adoption}). \\
			
			AI decision support
			& Tactical questions (pump operation, water rescue, hazmat data) were answered using hard-copy manuals or by radio; some feedback entries (11\,/\,204; Sect.~\ref{sec:feedback}) suggested retrieval-augmented support leveraging the operational manuals.
			& \texttt{fireops-ai-chat} (pre-production): retrieval-augmented assistant over the fire-service corpus, with per-incident transcripts. Retrieval-quality and hallucination-rate evaluation deferred to future work. \\
			
			Dual-ingestion redundancy
			& Single ingestion = single point of failure. Practitioners report the server feed degrades under regional peak load while radio-side WAS keeps delivering pages, arguing for deduplicated parallel ingest.
			& Parallel ingestion of radio (WAS) and server feeds with backend deduplication runs in production (Sect.~\ref{sec:architecture}); its resilience benefit is not yet quantitatively measured. \\
			
			Reach beyond the smartphone
			& Commanders relied on smartphones at the briefing table, constraining group awareness; members without the app received no parallel notification.
			& Permanent \texttt{edge-dashboard} on a station-house TV mirroring map and acknowledgement state, plus an \texttt{edge-notifier} that broadcasts alerts to per-brigade messaging-service channels (Telegram, WhatsApp, etc.). \\
			\hline
		\end{tabular}
	\end{table}
	
	The responses in Table~\ref{tab:reflection} shipped in successive releases over the 20~months after the flood: the acknowledgement layer, station-house dashboard and messaging-service bridge first, then the dual-ingestion path in 2025, the public iOS\,/\,Android release on 24~December~2025, and the pre-production AI assistant most recently.
	
	\section{Discussion}
	\label{sec:discussion}
	
	\subsection{Integration versus modularity}
	
	FireOPS integrates alerting, navigation, mapping and manuals into a single platform, reducing app-switching overhead during high-tempo response and raising a modularity question: if one module fails, should the others remain operational? The dual-ingestion redundancy of Sect.~\ref{sec:architecture} already addresses the most critical case, but a microservices refactor with standards-based open interfaces (CAP, OGC SensorThings) would decouple the remaining functions while preserving a unified user experience. The same boundary governs transferability beyond Austria: 112 is EU-wide, but alerting and dispatch infrastructure below the public-safety answering point is national. Everything above the ingestion layer is country-agnostic, so porting FireOPS is largely a matter of substituting the WAS, ELS-Export and syBOS adapters---which is a further argument for the standards-based refactor above.
	
	\subsection{Human-centric design and adoption}
	
	ISO~9241-210:2019~\cite{iso9241} characterises a human-centred system by understanding of users and tasks, user involvement throughout design, and evaluation-driven iteration. FireOPS was developed broadly consistently with these principles, though without formal usability evaluation: it was conceived by firefighters, refined using deployment feedback and aligned with the workflows of commanders and responders.

	\subsection{Limitations and future work}
	\label{sec:future_work}
	
	The present analysis is limited to a single district and a single anchor event, and provides descriptive baselines rather than controlled outcome evidence. A defensible follow-on evaluation should combine three streams: (1)~log-derived backend metrics (dispatch-to-acknowledgement latency, acknowledgement completeness, station-display update latency) segmented by incident type and time of day; (2)~semi-structured interviews with a stratified sample of incident commanders, machinists and rank-and-file responders, instrumented with an established acceptance framework~\cite{weidinger2024,weidinger2018}; and (3)~a before/after workflow comparison against the radio-and-paper baseline for at least one comparable high-load event. Beyond evaluation, future work will explore predictive resource allocation, alongside structured training and in-app onboarding to close the digital-literacy gap.
	
	\section{Conclusion}
	\label{sec:conclusion}
	
	FireOPS shows how volunteer fire brigades can integrate alerting, navigation, a shared COP and contextual knowledge into a single edge\,+\,cloud platform with a dual-ingestion (radio\,+\,server) redundancy path and a retrieval-augmented knowledge layer. Anchored in the flood-log extract, adoption analysis and 204-entry feedback corpus, this report identifies practitioner-led development and exposure to heterogeneous incident types as the principal enablers, and controlled outcome evidence and standards-based interoperability as the open challenges. Three tentative observations follow from the flood case. First, the load-bearing integration problem in volunteer-responder settings is the disciplined fusion of incumbent, proprietary feeds (syBOS, WAS, FCM\,/\,APNs, paging XML) into one human-centric interface; the hard parts are operational, not algorithmic. Second, dual-ingestion is a plausible resilience lever: both paths run in parallel in production today, though their relative contribution remains unquantified. Third, adoption appears temporally aligned with public-distribution and recognition events---correlational in a single-site dataset, not causal. Together these position practitioner-led information integration as a viable and underrepresented methodology for human-centric decision support in volunteer emergency services.
	
	\subsubsection*{Acknowledgements, disclosures and data availability.}
	We thank the District Fire Command Perg, especially Chief Eduard Paireder, the members of the participating brigades, and the OÖLFV for providing operational data. \textbf{Conflict of interest:} the authors are the developers and maintainers of FireOPS and members of the District Fire Command Perg; the platform is non-commercial and no revenue accrues to the authors. \textbf{Ethics:} only municipality- and unit-level aggregates of operational records are reported; no individuals are identifiable; feedback entries were submitted voluntarily via the app's feedback channel; processing complies with GDPR. \textbf{Use of AI:} Claude (Anthropic) was used for language editing and to assist the statistical analysis and figure generation reported in Sect.~\ref{sec:log_evidence}; the authors verified all outputs against the primary data and take full responsibility for the content. \textbf{Data:} the extract and registry snapshot underlying Fig.~\ref{fig:adoption} and Fig.~\ref{fig:flood_evidence} are available from the authors upon reasonable request.
	
	\bibliographystyle{splncs04}
	\bibliography{references}

@article{endsley1995,
  author  = {Endsley, Mica R.},
  title   = {{Toward a Theory of Situation Awareness in Dynamic Systems}},
  journal = {Human Factors},
  volume  = {37},
  number  = {1},
  pages   = {32--64},
  year    = {1995},
  doi     = {10.1518/001872095779049543}
}

@article{wolbers2013,
  author  = {Wolbers, Jeroen and Boersma, Kees},
  title   = {{The Common Operational Picture as Collective Sensemaking}},
  journal = {Journal of Contingencies and Crisis Management},
  volume  = {21},
  number  = {4},
  pages   = {186--199},
  year    = {2013},
  doi     = {10.1111/1468-5973.12027}
}

@article{steentveit2021,
  author  = {Steen-Tveit, Kristine and Munkvold, Bj{\o}rn Erik},
  title   = {{From Common Operational Picture to Common Situational Understanding: An Analysis Based on Practitioner Perspectives}},
  journal = {Safety Science},
  volume  = {142},
  pages   = {105381},
  year    = {2021},
  doi     = {10.1016/j.ssci.2021.105381}
}

@article{bharosa2010,
  author  = {Bharosa, Nitesh and Lee, JinKyu and Janssen, Marijn},
  title   = {{Challenges and Obstacles in Sharing and Coordinating Information During Multi-agency Disaster Response: Propositions from Field Exercises}},
  journal = {Information Systems Frontiers},
  volume  = {12},
  number  = {1},
  pages   = {49--65},
  year    = {2010},
  doi     = {10.1007/s10796-009-9174-z}
}

@article{yang2009,
  author  = {Yang, Lili and Prasanna, Raj and King, Malcolm},
  title   = {{On-site Information Systems Design for Emergency First Responders}},
  journal = {Journal of Information Technology Theory and Application (JITTA)},
  volume  = {10},
  number  = {1},
  pages   = {Article 2},
  year    = {2009}
}

@article{weidinger2018,
  author  = {Weidinger, Julian and Schlauderer, Sebastian and Overhage, Sven},
  title   = {{Is the Frontier Shifting into the Right Direction? A Qualitative Analysis of Acceptance Factors for Novel Firefighter Information Technologies}},
  journal = {Information Systems Frontiers},
  volume  = {20},
  number  = {4},
  pages   = {669--692},
  year    = {2018},
  doi     = {10.1007/s10796-017-9785-8}
}

@article{weidinger2024,
  author  = {Weidinger, Julian and Schlauderer, Sebastian and Overhage, Sven},
  title   = {{Determinants for the Acceptance of Emergency Response Information Systems: Ethnographical Insights into the Digitalization of a Voluntary Fire Department}},
  journal = {International Journal of Disaster Risk Reduction},
  volume  = {109},
  pages   = {104603},
  year    = {2024},
  doi     = {10.1016/j.ijdrr.2024.104603}
}

@inproceedings{kapalo2019,
  author    = {Kapalo, Katelynn A. and LaViola, Joseph J.},
  title     = {{Failing to Plan is Planning to Fail: Capturing the Pre-incident Planning Needs of Firefighters}},
  booktitle = {Proceedings of the Human Factors and Ergonomics Society Annual Meeting},
  volume    = {63},
  pages     = {612--616},
  year      = {2019},
  doi       = {10.1177/1071181319631113}
}

@article{kalabokidis2013,
  author  = {Kalabokidis, Kostas and Athanasis, Nikos and Gagliardi, Francesco and Karayiannis, Fotis and Palaiologou, Palaiologos and Parastatidis, Stavros and Vasilakos, Christos},
  title   = {{Virtual {Fire}: A Web-based {GIS} Platform for Forest Fire Control}},
  journal = {Ecological Informatics},
  volume  = {16},
  pages   = {62--69},
  year    = {2013},
  doi     = {10.1016/j.ecoinf.2013.04.007}
}

@article{noonanwright2011,
  author  = {Noonan-Wright, Erin K. and Opperman, Tonja S. and Finney, Mark A.
             and Zimmerman, G. Thomas and Seli, Robert C. and Elenz, Lisa M.
             and Calkin, David E. and Fiedler, John R.},
  title   = {{Developing the US Wildland Fire Decision Support System}},
  journal = {Journal of Combustion},
  volume  = {2011},
  pages   = {168473},
  year    = {2011},
  doi     = {10.1155/2011/168473}
}

@article{kaufhold2024,
  author  = {Kaufhold, Marc-Andr{\'e}},
  title   = {{Exploring the Evolving Landscape of Human-centred Crisis Informatics: Current Challenges and Future Trends}},
  journal = {i-com},
  volume  = {23},
  number  = {2},
  pages   = {155--163},
  year    = {2024},
  doi     = {10.1515/icom-2024-0002}
}

@incollection{steentveit2023,
  author    = {Steen-Tveit, Kristine and Munkvold, Bj{\o}rn Erik},
  title     = {{Defining Common Information Requirements for Supporting Multiagency Emergency Operations}},
  booktitle = {Disaster Management and Information Technology},
  editor    = {Scholl, Hans Jochen and Holdeman, Eric E. and Boersma, F. Kees},
  series    = {Public Administration and Information Technology},
  volume    = {40},
  pages     = {307--322},
  publisher = {Springer},
  year      = {2023},
  doi       = {10.1007/978-3-031-20939-0_14}
}

@inproceedings{nunavath2019,
  author    = {Nunavath, Vimala and Prinz, Andreas},
  title     = {{{LifeRescue}: A Web-based Application for Emergency Responders During Fire Emergency Response}},
  booktitle = {Proc.\ 3rd International Conference on Information and Communication Technologies for Disaster Management (ICT-DM)},
  publisher = {IEEE},
  year      = {2016},
  doi       = {10.1109/ICT-DM.2016.7857204}
}

@misc{orf2024,
  author       = {{ORF Ober\"osterreich}},
  title        = {{Noch keine Hochwasser-Entwarnung in {O\"O}}},
  year         = {2024},
  url          = {https://ooe.orf.at/stories/3273098/}
}

@misc{fireops_pe_ooelfv,
  author       = {{Bezirksfeuerwehrkommando Perg}},
  title        = {{{FireOPS} - Einsatzf\"uhrung \& Lagemanagement}},
  howpublished = {Project page, OÖ Landesfeuerwehrverband},
  url          = {https://pe.ooelfv.at/fireops},
  year         = {2024}
}

@misc{digitalos2025,
  author       = {Pammer, Anja},
  title        = {{FireOPS triumphiert beim Digitalos~2025 in der Kategorie \enquote{Digitales Projekt}}},
  year         = {2025},
  url          = {https://pe.ooelfv.at/aktuelles/news/artikel/fireops-triumphiert-beim-digitalos-2025-in-der-kategorie-digitales-projekt}
}

@misc{regionalitaetspreis2025,
  author       = {{MeinBezirk Perg}},
  title        = {{Regionalit\"atspreis~2025: Sonderpreis \enquote{Regional \& Digital} f\"ur {FireOPS}}},
  year         = {2025},
  url          = {https://www.meinbezirk.at/oberoesterreich/c-lokales/sonderpreis-regional-und-digital-fuer-fireops_a7589779}
}

@misc{danubecup2026,
  author       = {K{\"o}ck, Michael},
  title        = {{Einsatzplattform \enquote{FireOPS} steht im Finale des Danube Cup~2026}},
  year         = {2026},
  url          = {https://www.meinbezirk.at/perg/c-lokales/fireops-steht-im-finale-des-danube-cup-2026_a8631432}
}

@misc{umweltpreis2024,
  author       = {{MeinBezirk Perg}},
  title        = {{Umweltpreis O\"O 2024: Revolution\"are Einsatzf\"uhrung ausgezeichnet}},
  year         = {2024},
  url          = {https://www.meinbezirk.at/perg/c-lokales/revolutionaere-einsatzfuehrung-ausgezeichnet_a6992163}
}

@misc{divera247,
  author       = {{DIVERA GmbH}},
  title        = {{{DIVERA 24/7} Alarmierungs- und Verf\"ugbarkeitssystem - Produktbeschreibung}},
  year         = {2026},
  url          = {https://www.divera247.com/versionen/free.html}
}

@misc{blaulichtsms,
  author       = {{blaulichtSMS GmbH}},
  title        = {{blaulicht{SMS} - Alle Features im \"Uberblick}},
  year         = {2026},
  url          = {https://blaulichtsms.net/features/}
}

@misc{landooe2024,
  author       = {{Amt der O\"O Landesregierung, Abt.\ Wasserwirtschaft}},
  title        = {{Analyse des Hochwasserereignisses im September 2024 hervorgerufen durch eine Vb-Wetterlage - Kurzfassung}},
  year         = {2024},
  url          = {https://www.land-oberoesterreich.gv.at/Mediendateien/Formulare/Dokumente\%20UWD\%20Abt_WW/Analyse\%20des\%20Hochwasserereignisses_2024_Kurzfassung_ohne\%20Anha.pdf}
}

@misc{oasis_cap,
  author       = {{OASIS}},
  title        = {{Common Alerting Protocol ({CAP}) Version 1.2}},
  howpublished = {OASIS Standard},
  year         = {2010}
}

@misc{ogc_sensorthings,
  author       = {{Open Geospatial Consortium}},
  title        = {{{SensorThings API} Part 1: Sensing, Version 1.1}},
  howpublished = {OGC 18-088},
  year         = {2021}
}

@misc{iso19168,
  author       = {{International Organization for Standardization}},
  title        = {{{ISO} 19168-1:2020, Geographic Information - Geospatial {API} for Features - Part 1: Core ({OGC API} - Features)}},
  year         = {2020}
}

@misc{iso9241,
  author       = {{International Organization for Standardization}},
  title        = {{{ISO} 9241-210:2019, Ergonomics of Human-system Interaction - Part 210: Human-centred Design for Interactive Systems}},
  year         = {2019}
}

@misc{strabag_alu2g,
  author       = {{STRABAG Infrastructure \& Safety Solutions GmbH}},
  title        = {{Digital Alarm Repeater and Terminals}},
  year         = {2026},
  url          = {https://www.strabag-iss.com/en/services/digital-alerting/digital-alarm-repeater-and-terminals}
}

@misc{oebfv2024,
  author       = {{{\"O}sterreichischer Bundesfeuerwehrverband}},
  title        = {{Feuerwehr in {\"O}sterreich 2024 und im Verlauf 2008 bis 2024}},
  howpublished = {Statistics sheet, {\"O}BFV},
  year         = {2025},
  url          = {https://www.bundesfeuerwehrverband.at/wp-content/uploads/2025/02/Statistik_2024.pdf},
  urldate      = {2026-07-31}
}

@misc{rabbitmq,
  author       = {{Broadcom}},
  title        = {{RabbitMQ: One broker to queue them all}},
  year         = {2026},
  howpublished = {\url{https://www.rabbitmq.com/}},
  urldate = {2026-07-03}
}

@misc{bmi_warnung_alarmierung,
  author  = {{Bundesministerium für Inneres}},
  title   = {{Warnung und Alarmierung}},
  year    = {2024},
  url     = {https://www.bmi.gv.at/204/skkm/warnung.html},
  urldate = {2026-07-03}
}

@misc{solarys_sybos,
  author  = {{SOLARYS Software GmbH}},
  title   = {{syBOS: Die umfassende Verwaltungssoftware für Einsatzorganisationen}},
  year    = {2026},
  url     = {https://solarys.com/verwaltung/},
  urldate = {2026-07-03}
}
\end{document}